\documentclass[a4paper,twocolumn,aps,prb,superscriptaddress,showpacs]{revtex4}
\usepackage{dcolumn}
\usepackage{amsmath}
\usepackage{graphicx}
\usepackage{latexsym}  
\usepackage{amsfonts}  
\usepackage{amssymb}
\usepackage{color}  
\usepackage{bbm}
\usepackage{graphicx}
\DeclareGraphicsExtensions{.eps,.pdf,.png,.gif,.jpg}

\newcommand{\be}{\begin{equation}}
\newcommand{\beq}{\begin{equation}}
\newcommand{\ee}{\end{equation}}
\newcommand{\bea}{\begin{eqnarray}}
\newcommand{\eea}{\end{eqnarray}}
\newcommand{\ba}{\begin{array}}
\newcommand{\ea}{\end{array}}

\newcommand{\Iid}[1] {\int \! \! {\rm d} #1}

\newcommand{\G}{\Gamma}

\newcommand{\vG} {{\bf G}}

\newcommand{\nn} {\nonumber}

\newcounter{saveeqn}

\def\a{\alpha}

\def\ve{\varepsilon}
\def\s{\sigma}
\def\ra{\rightarrow}

\newcommand{\te}{\textrm}

\begin{document}

\title{Comparative study of many-body perturbation theory and 
time-dependent density functional theory in the out-of-equilibrium
Anderson model}

\author{A.-M.~Uimonen}
\affiliation{ Department of Physics, Nanoscience Center, FI 40014, 
University of Jyv\"askyl\"a, Jyv\"askyl\"a, Finland}
\affiliation{European Theoretical Spectroscopy Facility (ETSF)}

\author{E.~Khosravi}
\affiliation{Max-Planck Institut f\"ur Mikrostrukturphysik, Weinberg 2, 
D-06120 Halle, Germany}
\affiliation{European Theoretical Spectroscopy Facility (ETSF)}

\author{A.~Stan}
\affiliation{ Department of Physics, Nanoscience Center, FI 40014, 
University of Jyv\"askyl\"a, Jyv\"askyl\"a, Finland}
\affiliation{European Theoretical Spectroscopy Facility (ETSF)}

\author{G.~Stefanucci}
\affiliation{Dipartimento di Fisica, Universit\`{a} di Roma Tor Vergata,
Via della Ricerca Scientifica 1, 00133 Rome, Italy}
\affiliation{INFN, Laboratori Nazionali di Frascati, Via E. Fermi 40, 00044 Frascati, Italy}
\affiliation{European Theoretical Spectroscopy Facility (ETSF)}

\author{S.~Kurth}  
\affiliation{Nano-Bio Spectroscopy Group, 
Departamento de F\'{i}sica de Materiales, 
Universidad del Pa\'{i}s Vasco UPV/EHU, Centro F\'{i}sica de Materiales 
CSIC-UPV/EHU, Avenida de Tolosa 72, E-20018 San Sebasti\'{a}n, Spain} 
\affiliation{IKERBASQUE, Basque Foundation for Science, E-48011 Bilbao, Spain}
\affiliation{European Theoretical Spectroscopy Facility (ETSF)}

\author{R.~van Leeuwen}
\affiliation{ Department of Physics, Nanoscience Center, FI 40014, 
University of Jyv\"askyl\"a, Jyv\"askyl\"a, Finland}
\affiliation{European Theoretical Spectroscopy Facility (ETSF)}

\author{E. K. U.~Gross}
\affiliation{Max-Planck Institut f\"ur Mikrostrukturphysik, Weinberg 2, 
D-06120 Halle, Germany}
\affiliation{European Theoretical Spectroscopy Facility (ETSF)}

\date{\today}  

\begin{abstract}
We study time-dependent electron transport through an Anderson model. 
The electronic interactions on the impurity site are included via the self-energy approximations 
at Hartree-Fock (HF), second Born (2B), GW, and T-Matrix level as well as within a time-dependent 
density functional (TDDFT) scheme based on the adiabatic Bethe-Ansatz local density approximation (ABALDA) for the exchange correlation 
potential. The Anderson model is driven out of equilibrium by applying a bias to the leads
and its nonequilibrium dynamics is determined by real-time propagation. The time-dependent currents and densities 
are compared to benchmark results obtained with the time-dependent density matrix 
renormalization group (tDMRG) method. Many-body perturbation theory beyond 
HF gives results in close agreement with tDMRG especially within the 2B approximation.
We find that the TDDFT approach with the ABALDA approximation produces accurate results for the
densities on the impurity site but overestimates the currents. This problem is
found to have its origin in an overestimation of the lead densities which indicates 
that the exchange correlation potential must attain nonzero values in the leads. 
\end{abstract}

  
\maketitle  

\section{Introduction}

The process of electron transport through molecules and nanostructures is part of a rapidly 
growing research area in condensed matter physics~\cite{Datta_book, DiVentra_book}. 
On a fundamental level one has to deal with time-dependent processes in an open system 
where different scattering mechanisms such as electron-electron or electron-phonon interactions
are of great importance. These factors make the transport problem not only difficult, but also very rich in 
physical phenomena. Most of the recent studies in molecular electronics have focused on the 
description of steady-state transport while neglecting short-time dynamics such as transients 
and fast switching processes. However, these processes will become increasingly important 
since fast switching rates play a pivotal role in the operation of future devices. 

For the description of electron transport several numerical 
approaches have been developed that can deal with fully time-dependent systems~\cite{Gurviz:96,StefanucciAlmbladh:04,Moldoveanu:07,Bokes:08,Ryndyk:08,MyohanenStanStefanucciLeeuwen:09,Andergassen:10,Zheng:10,Zheng:07,FriesenVerdozziAlmbladh:09, FriesenVerdozziAlmbladh-2, Brandschaedel:10, KurthStefanucciAlmbladhRubioGross:05}.  
Among these are the time-dependent density matrix renormalization group (tDMRG) approach~\cite{Boulat:08},
time-dependent density functional theory (TDDFT)~\cite{StefanucciAlmbladh:04, KurthStefanucciAlmbladhRubioGross:05}, and self-consistent 
many-body perturbation theory (MBPT) based on the Kadanoff-Baym (KB) equations~\cite{MyohanenStanStefanucciLeeuwen:08,FriesenVerdozziAlmbladh:09, FriesenVerdozziAlmbladh-2}.
Each of these methods has its own advantages and disadvantages. To the best
of our knowledge, a comparative study of these three approaches on an identical 
time-dependent system has not been carried out. Such a study would be very valuable for gaining insight 
into these methods and into the 
direction in which each method needs to be improved.
In the steady-state regime of quantum transport such comparisons of many-body and benchmark approaches
were made by Wang {\em et al.} \cite{WangSpataruHybertsenMillis:08} within
the GW-approximation and by Schmitt and Anders~\cite{SchmittAnders:10} at second Born (2B) 
and GW level. In both cases good agreement with benchmark results was found in certain parameter ranges.
We want to extend these comparisons to the transient regime as well.

Let us give a brief description of the approaches that we use in this work.
The tDMRG method is a numerical algorithm based on truncation of the Hilbert space of low-dimensional systems~\cite{schollwock:04,KarraschSchoeller:10,KarraschMeden:10,MetzerSchoenhammer:11}.  
In this work we did not carry out such calculations ourselves but we use published tDMRG 
results\cite{HeidrichMeisnerFeiguinDagotto:09}  as a benchmark for both the TDDFT and MBPT approaches.

In the TDDFT approach \cite{Gross_book,TDDFT_book} a system of interacting electrons is mapped, in an
exact manner, onto a system of noninteracting electrons moving in an effective
time-dependent external potential known as the Kohn-Sham (KS) potential.
The KS potential is functionally dependent on the electron density such 
that it produces a KS wave function with a density identical to the
time-dependent density of the interacting system. It is important to note that TDDFT yields 
in principle the exact time-dependent current through a molecular junction~\cite{StefanucciAlmbladh:04,DiventraTodorov:04}.  
The use of one-particle equations in TDDFT allows for large scale first-principle 
calculations on realistic systems. 
In practice, however, approximations are unavoidable and the accuracy of a TDDFT
calculation crucially depends on the quality of the approximate exchange-correlation ($\textrm{\small XC}$) potential used.
Most applications of TDDFT to quantum transport 
processes\cite{EversWeigendKoentopp:04,BaerSeidemanIlaniNeuhauser:04,StefanucciAlmbladh:04,StefanucciAlmbladh:04-2,KurthStefanucciAlmbladhRubioGross:05,BurkeCarGebauer:05,Saietal:07} use the adiabatic approximation which assumes that 
the $\textrm{\small XC}$-potential instantaneously follows the density profile. This is a reasonable assumption
when the density changes are slow on a time-scale of typical lead-to-molecule tunneling 
rates, and also when the switch-on times of the applied biases are small enough.
However, it has also been pointed out that non-adiabatic effects can have substantial 
influence~\cite{Saietal:05,VignaleDiventra:09} on calculated properties.
In such cases there is also a need to introduce spatial non-locality in the density
functional because the non-localities in space and time are strongly related by conservation laws \cite{VignaleKohn:96}.
This relation is virtually unexplored within a quantum transport context. Gaining further insight into this 
issue is one of the goals of this work. 

The MBPT approach based on the KB equations\cite{KadanoffBaym:62,Danielewicz:84} 
has been successfully 
applied to time-dependent quantum transport for model 
systems\cite{MyohanenStanStefanucciLeeuwen:08,MyohanenStanStefanucciLeeuwen:09, FriesenVerdozziAlmbladh:09, FriesenVerdozziAlmbladh-2}.
The method offers the possibility of including relevant physical processes by means of selection of 
Feynman diagrams for the self-energy. The electron-electron correlations are thus considered 
via the many-body self-energy term which is treated perturbatively to infinite order by summation
of infinite classes of diagrams. 
Furthermore by using conserving approximations\cite{BaymKadanoff:61,Baym:62}
such as the Hartree-Fock (HF), second Born (2B), GW, and T-Matrix approximations we can
guarantee that conservation laws are obeyed, which has shown to be very important 
in quantum transport\cite{ThygesenRubio:08, Strange:11}.
In this approach one has direct access to quantities like quasiparticle spectra, lifetimes, and 
screened interactions which provide insight into the effects of electron correlation. 
In particular, the non-locality in time of the 2B, GW, and T-Matrix approximations allows 
for a description of memory effects and quasi-particle broadening. 
We use the partition-free scheme where the device is initially contacted to the leads 
and the whole system in thermal equilibrium \cite{Cini:80}. 
In this approach both the transient 
and steady-state currents have a direct physical meaning as these currents are
induced by the physical switch-on of a bias. In the partitioned approaches
they are instead induced by switch-on of a device-lead coupling which does not
correspond to the standard experimental situation. We finally like to point out that
the MBPT approach can be used to derive new improved time-dependent density functionals 
with memory and conserving properties~\cite{vonBarthetal2005}.
This has been done successfully within the linear response regime~\cite{Hellgren,Hellgren:10}.

Since both TDDFT and MBPT require the use of approximations it is important 
to have independent benchmark results. For the Anderson impurity model such
results in the time domain have recently been obtained with tDMRG \cite{HeidrichMeisnerFeiguinDagotto:09}. 
Therefore we will use this system as a test case for our comparative study of MBPT and TDDFT.
The paper is organized as follows: 
in Sec.~\ref{model} we introduce the model used in our investigation. In 
Sections~\ref{sec:mbpt} and~\ref{tddftsec} we describe the MBPT and the TDDFT methods used. 
In Sec.~\ref{results_weak} we present numerical 
results and the last section summarizes our conclusions.

\section{The model}
\label{model}
We study an Anderson impurity model\cite{Anderson:61} described by
the Hamiltonian
\be
\hat{H}(t)=\hat{H}_{\textrm{\tiny C}}+ \sum_{\alpha} \hat{H}_\alpha (t) + \hat{H}_{\rm{\tiny T}},
\label{modelHam}
\ee
where $\hat{H}_{\rm{\tiny C}}$, $\hat{H}_{\a}$, and $\hat{H}_{\textrm{\tiny T}}$ respectively describe the impurity region,
the leads $\alpha$ {\rm{\small (= L,R)}}, and the 
tunneling between the impurity region and the leads.
The Hamiltonian for the impurity site reads
\begin{equation}
\hat{H}_{\textrm{\tiny C}} = \sum_{\sigma} \varepsilon_{0} 
\hat{c}_{0\sigma}^{\dagger} \hat{c}_{0\sigma} + \frac{1}{2} \sum_{\sigma,\sigma'}
U\hat{c}_{0\sigma}^{\dagger} \hat{c}_{0\sigma'}^{\dagger} 
\hat{c}_{0\sigma'} \hat{c}_{0\sigma},
\end{equation}
where $c_{\sigma}^{\dagger},c_{\sigma}$ are fermionic creation and annihilation operators
and ${\sigma, \sigma'}$ are the spin indices, $\varepsilon_{0}$ is 
the on-site energy of the interacting site and $U$ is the interaction term or the charging energy. 
The Hamiltonian $\hat{H}_{\a}(t)$, describing the leads is 
\begin{eqnarray}
\hat{H}_\alpha (t) &=&   \sum_{\substack{ \s}}\sum_{\substack{i=1}}^{\infty} \Big( \varepsilon_{\alpha} + 
W_{\alpha}(t) \Big) 
\hat{c}_{i\sigma\alpha}^{\dagger} \hat{c}_{i\sigma\alpha} \nn \\
&&  - \sum_{\substack{ \s}} \sum_{\substack{ i=1}}^{\infty} \Big( V_{\alpha} \hat{c}_{i\sigma\alpha}^{\dagger} 
\hat{c}_{i+1\sigma\alpha} + H.c. \Big),
\end{eqnarray}
where $\varepsilon_{\alpha}$ is the on-site energy in the leads,
$W_{\alpha}$ is the bias on the lead $\alpha$ and $V_{\alpha}$ is the hopping 
between neighboring lead sites. The tunneling Hamiltonian 
describes the coupling between the impurity site and the leads, and has the form
\be
\hat{H}_{\textrm{\tiny T}} = - \sum_{\sigma} \left( 
V_{\textrm{link}}\, \hat{c}_{0\sigma}^{\dagger} \hat{c}_{1\sigma L} 
+ V_{\textrm{link}}\, \hat{c}_{0 \sigma}^{\dagger} \hat{c}_{1\sigma R}  
+ H.c. \right), 
\ee
where $V_{\textrm{link}}$ is the hopping from the leads to the impurity site
and vice versa.

\subsection{Kadanoff-Baym equations}
\label{sec:mbpt}

The nonequilibrium properties of the system are studied with the aid of nonequilibrium Green
function theory and TDDFT described later in section~\ref{tddftsec}. 
The nonequilibrium Green function is defined as
the expectation value with respect to the initial state of the contour-ordered product of
creation and annihilation operators~\cite{Danielewicz:84}
\begin{eqnarray}
\label{green_def}
G_{i\sigma, j\sigma'}(z,z')&=&-i\big\langle\mathcal{T}[\hat{c}_{H,i\sigma}(z) \hat{c}_{H,j\sigma'}^{\dagger}(z')] \big\rangle,
\end{eqnarray}
where $i,j$ are the site-indices, $\mathcal{T}$ denotes the time-ordering operator along the 
Keldysh contour~\cite{Danielewicz:84}, and where the contour variables $z$ and $z'$ specify the 
position on the contour~\cite{intro_Keldysh:08}. 
The subscript $H$ refers to operators in the Heisenberg picture with respect to the time-dependent Hamiltonian $\hat{H}(z)$~\cite{Danielewicz:84,intro_Keldysh:08}.
The Green function of the whole system satisfies the equation of motion 
\begin{eqnarray}
\label{EOM}
[i\partial_{z}\mathbf{1} &-&\mathbf{H}(z)]\vG(z,z')=\delta(z,z')\mathbf{1} + \nonumber \\
&+&\int_{\mathcal{C}}\textrm{d}\bar{z}\;\mathbf{\Sigma}^{\textrm{\tiny MB}}[\vG](z,\bar{z}) \vG(\bar{z},z'),
\end{eqnarray}
where we introduced the many-body self-energy $\Sigma^{\textrm{\tiny MB}} [G]$ 
which accounts for all the exchange and correlation effects~\cite{MyohanenStanStefanucciLeeuwen:09} 
and where we suppressed spatial indices. The self-energy is a functional of the Green function which in practice is defined
diagrammatically~\cite{Danielewicz:84,KadanoffBaym:62}. 

\begin{figure}[t]
	\begin{center}
		\includegraphics[width=0.49\textwidth]{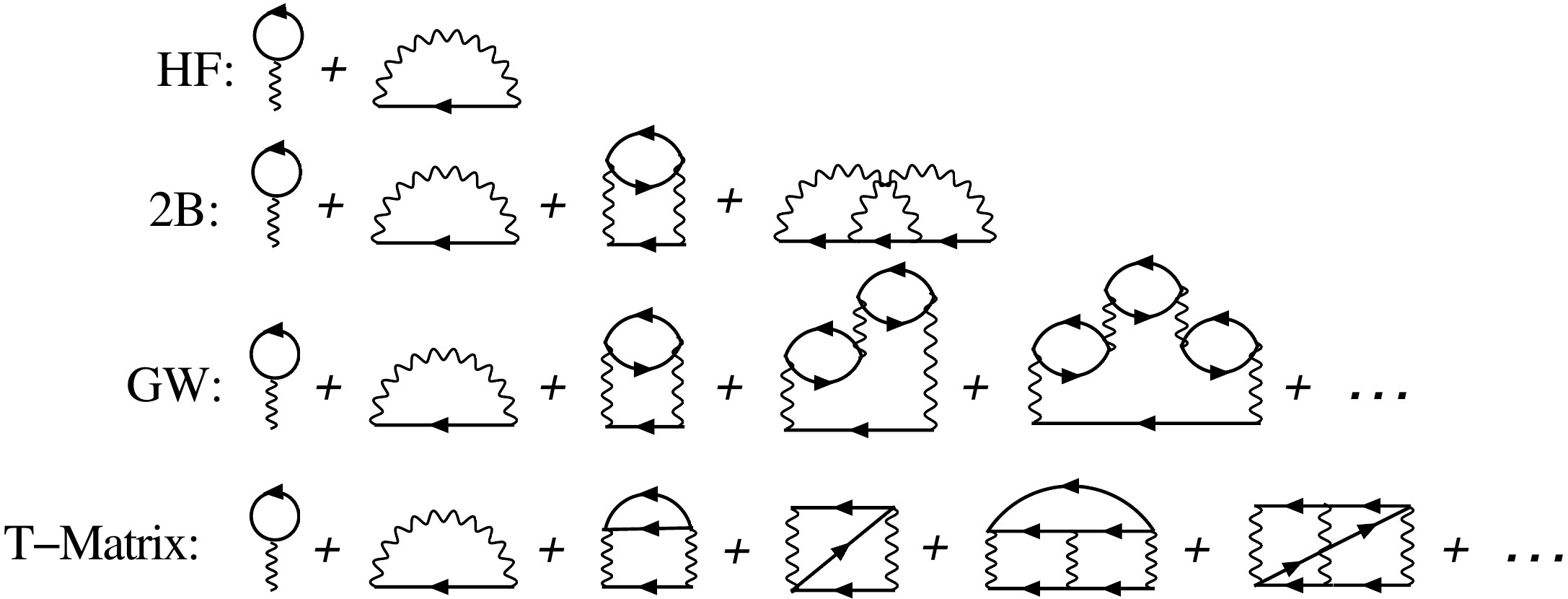}
		\caption{Diagrammatic representation of the conserving many-body approximations to the self-energy. 
		 Wiggly lines denote the many-body interaction. All Green function lines (directed solid lines) are
		are fully dressed.}
	\label{diagrams}
	\end{center}
\end{figure}

In this work we solve the equation of motion of 
the Keldysh Green function fully self-consistently~\cite{dahlen06procKB, DahlenLeeuwen:07, sdvl.2009method, sdvl.2009gw}
using the four approximations of the many-body self-energy $\Sigma^{\textrm{\tiny{MB}}}[G]$ shown in Fig.~\ref{diagrams}.
The self-consistent HF approximation is time-local and includes 
the Hartree and the exchange potential.
The self-consistent 2B approximation consists of the two diagrams to second order in the
interaction~\cite{DahlenLeeuwen:05}. It describes dynamical screening of the electron-electron 
interaction via a simple bubble diagram and includes a vertex contribution via the second order exchange diagram.
The fully self-consistent GW approximation~\cite{Hedin} incorporates the dynamical screening effects via the 
infinite summation of bubble diagrams~\cite{sdvl.2009gw}. In this approximation, the Coulomb interaction
is replaced by the screened potential $W$. The last approximation we use is the fully self-consistent 
T-matrix approximation~\cite{KadanoffBaym:62, Danielewicz:84}. 
It contains the 2B diagrams and an infinite summation of the ladder diagrams. The GW and T-matrix approximations
are complementary since the GW approximation 
accounts for dynamical screening in infinite systems with long-range Coulombic interactions whereas the T-matrix
approximation is known to be important in describing infinite systems with a short range hard-core interaction \cite{KadanoffBaym:62,fetter}.

When we describe a system attached to noninteracting leads the equation of motion of the Green function for 
the whole system can be folded into an effective equation of motion of the Green function for the 
central region~\cite{MyohanenStanStefanucciLeeuwen:09,MyohanenStanStefanucciLeeuwen:08}.
In the case of the impurity model that we consider this gives
\begin{eqnarray}
\label{impurity_EOM}
&&\left[i\partial_z-H(z)\right]G(z,z') =\delta(z,z')\\
&&+\int_{\mathcal{C}}\,\textrm{d}\bar{z}\,\Big\{\big[\Sigma_{\textrm{em}}(z,\bar{z})+\Sigma^{\textrm{\tiny MB}}[G](z,\bar{z})\big]G(\bar{z},z')\Big\},\nonumber
\end{eqnarray}
where the embedding self-energy $\Sigma_{\textrm{em}}(z,z')$  accounts for the tunneling of electrons between 
leads and the impurity site. The many-body self-energy depends only on the Green function of the central site
as the many-body interaction is restricted to the central site only. This Green function has only one spatial index.

For the time-dependent observables calculated on the real axis we denote the
contour parameter $z$ by the real time $t$. The time-dependent density for the 
impurity site is given by
\begin{equation}
n_0(t) =-i \,G^<(t,t^+),
\end{equation}
where  $t^+$  approaches $t$ from an infinitesimally later time $t^+ = t+\delta$. 
The current through the lead $\alpha=$ {\rm{\small (L,R)}} can be expressed in terms of the so-called 
Keldysh Green functions as~\cite{MyohanenStanStefanucciLeeuwen:09,MyohanenStanStefanucciLeeuwen:08}
\begin{eqnarray}
\label{eq:current}
\textrm{I}_{\alpha}(t)&=&2\te{Re}\bigg\lbrace  \int_{t_0}^{t}d\bar{t}
[G^<(t,\bar{t})\Sigma_{\rm em,\alpha}^{\textrm{\tiny A}}(\bar{t},t)\nonumber\\
&+&\int_{t_0}^{t}d\bar{t}G^{\textrm{\tiny R}}(t,\bar{t})\Sigma_{\rm em,\alpha}^<(\bar{t},t)]\nonumber\\
&-i&\int_0^{\beta}d\bar{\tau}G^{\rceil}(t,\bar{\tau})\Sigma_{\rm em,\alpha}^{\lceil}(\bar{\tau},t)\bigg\rbrace,
\end{eqnarray}
where we integrated on the Keldysh contour and where the superscripts {\small $\rm A,R$} and {\small $\rm<$} refer 
to the advanced, retarded, and lesser 
components of the Green function and the self-energy. Further, {\footnotesize $\rceil$} and {\footnotesize $\lceil$} are the mixed components having one time 
argument on the imaginary axis and another on the  real axis \cite{sdvl.2009method, MyohanenStanStefanucciLeeuwen:09}.  
The initial many-body correlations and embedding 
effects are taken into account by the last term in  
equation~(\ref{eq:current}) which is an integral over the vertical track of the Keldysh contour~\cite{intro_Keldysh:08}.
If we assume that in the $t\to\infty$ limit  the terms with components on the imaginary track vanish 
and that the Green function and the self-energy depend only on $t-t'$ then we can
Fourier transform (\ref{eq:current}) and
we obtain the Meir-Wingreen formula for the steady-state current~\cite{MeirWingreen:92} 
\begin{equation}
\label{MW}
I_{\alpha}^{\infty}=-i\int_{-\infty}^{\infty}\frac{d\omega}{\pi}\G_{\alpha}(\omega)\Big[G^<(\omega)-2i\pi f_{\alpha}(\omega)A(\omega)\Big]
\end{equation}
where $\G_{\alpha}(\omega)$ is the imaginary part of the embedding self-energy, $f_{\alpha}(\omega)$ is the Fermi function 
and $A(\omega)$ is the steady-state spectral function~\cite{MeirWingreen:92}.
Hence, equation~(\ref{eq:current}) is a generalization of the Meir-Wingreen formula~\cite{MyohanenStanStefanucciLeeuwen:09}.\\
We further define the nonequilibrium spectral function 
\begin{equation}
\label{eq:spectral}
A(T,\omega)=-\textrm{Im}\int\frac{\textrm{d}\tau }{\pi } e^{i\omega\tau}\big[ G^>-G^<\big](T+\frac{\tau}{2},T-\frac{\tau}{2}),
\end{equation}
where $\tau=t-t'$ is a relative time and $T=(t+t')/2$ is an average 
time-coordinate~\cite{dahlen06procKB2, MyohanenStanStefanucciLeeuwen:09, uimonen:10}.
In equilibrium, this function is independent of $T$ and  
has peaks below the Fermi level at the electron removal energies of the system, 
while above the Fermi level it has peaks at the electron addition
energies. 
If the time-dependent external field becomes constant after some switching time, then also
the spectral function becomes
independent of $T$ after some transient period 
and has peaks at the addition and removal energies of the biased system~\cite{MyohanenStanStefanucciLeeuwen:10}.

\subsection{Time-dependent density functional theory}
\label{tddftsec}

\begin{figure}[b]
    \begin{center}
	\includegraphics[width=0.49\textwidth]{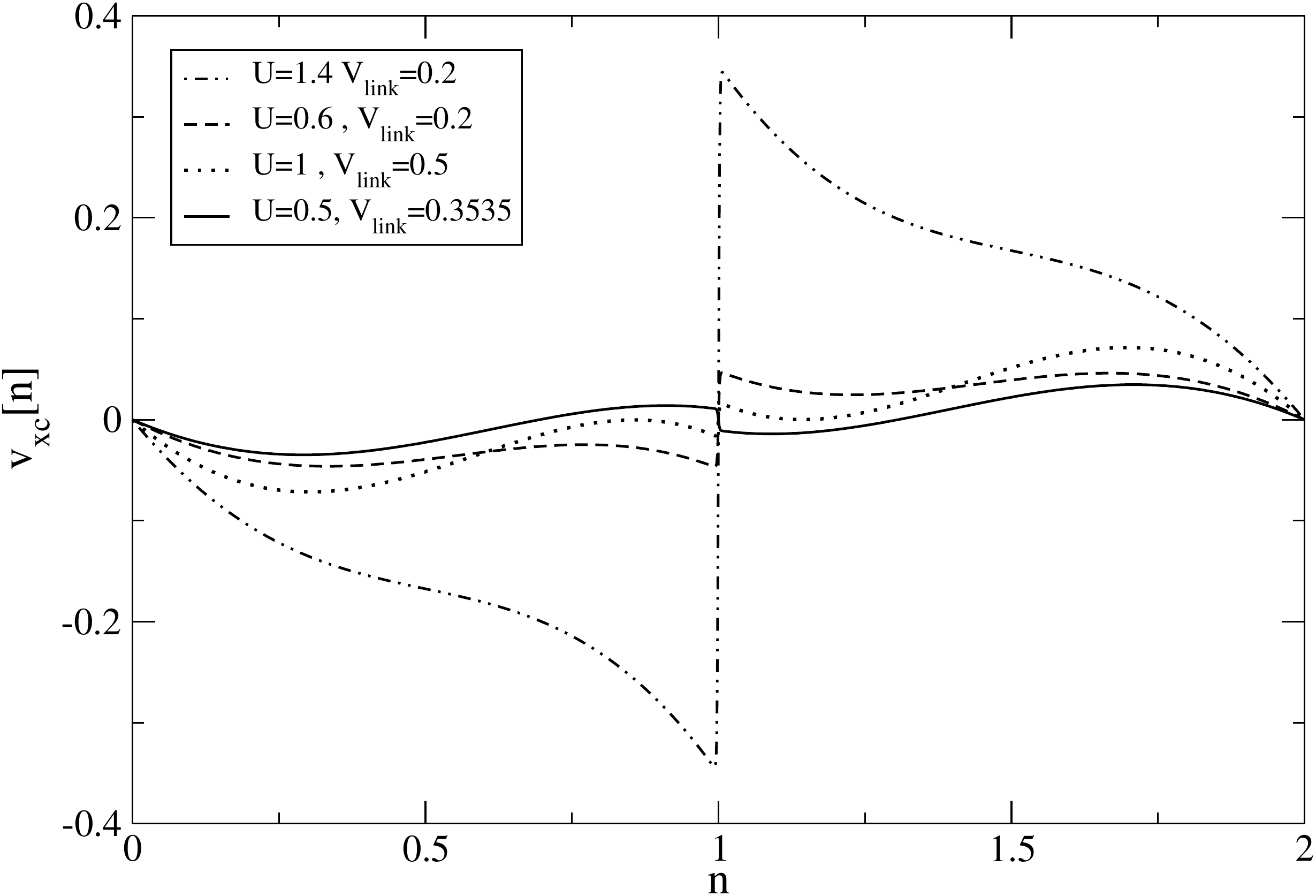}
	\caption{The BALDA $\textrm{\small XC}$-potential as a function of the density for	 
	parameters used in the subsequent sections.}
	\label{vxcplot}
\end{center}
\end{figure}

Within TDDFT the complication brought forward by considering an open 
system can be resolved in a very similar manner as in MBPT (see Section~\ref{sec:mbpt}),
with the aid of an embedding self-energy. 
The equation of motion for the $k$-th single-particle 
orbital is projected onto the Anderson impurity site and reads
\bea
\left[ i \partial_t - H^{\textrm{\tiny KS}}(t) \right] \psi_{k}(t) = 
\int_0^t {\rm d} \bar{t} \; \Sigma_{\rm KS}^{\textrm{\tiny R}}(t,\bar{t}) \psi_{k}(\bar{t}) 
\nn\\
+ \sum_{\alpha} V_{\textrm{link}}\, g_{\alpha \alpha}^{\textrm{\tiny R}}(t,0)
\psi_{k,\alpha}(0), 
\eea
where $\Sigma_{\rm KS}^{\textrm{\tiny R}}(t,\bar{t})$ is the KS embedding self-energy
and $g_{\alpha\alpha}^{\textrm{\tiny R}}$ is the retarded lead Green 
function. 
This expression is, in principle, exact. If we now assume that the exchange-correlation potential is
zero in the leads then $\Sigma_{\rm KS}^{\textrm{\tiny R}}(t,\bar{t})$ can be replaced
by $\Sigma_{\rm  em}^{\textrm{\tiny R}}(t,\bar{t})$ of Eq.(\ref{EOM}). 
We will assume this in the following. Then
for the Anderson impurity model the
KS Hamiltonian $H^{\textrm{\tiny KS}}(t)$ has the 
following from
\bea
H^{\textrm{\tiny KS}}(t) &=& v_{\rm KS}(t) 
\nn \\ &=& \varepsilon_0(t) + \frac{1}{2} U 
n_0(t) + v_{\rm xc}[n](t).
\label{kspot}
\eea
The approximation for the $\textrm{\small XC}$-potential in this work is based
on the local density approximation (LDA) for the static, non-uniform 
one-dimensional Hubbard model derived from the Bethe ansatz (Bethe ansatz LDA, 
BALDA) which has been suggested in Ref. [\onlinecite{schonhammerGunnarssonNoack:95}] and
further been developed in Ref. [\onlinecite{LimaSilvaOliveiraCapelle:03}]. The adiabatic version 
\cite{Verdozzi:08} of this functional (ABALDA) makes $v_{\rm xc}[n]$ local in 
both space and time. The modified version of ABALDA for the transport setup~\cite{KurthStefanucciKhosraviVerdozziGross:10} 
is taking into account the different hopping between the impurity site and the leads 
and reads explicitly
\be
v^{\textrm{BALDA}}_{\rm xc}[n] = \theta(1-n) v_{\rm xc}^{<}(n) - \theta(n-1) v_{\rm xc}^{<}(2-n) 
\label{abalda1},
\ee
where
\be
v_{\rm xc}^{<}(n) = - \frac{1}{2} U n - 2\, V_{\textrm{link}} \left[ \cos\left( 
\frac{\pi n}{2} \right) - \cos\left(\frac{\pi n}{\xi}\right) \right].
\label{abalda2}
\ee
Here, $\xi$ is a parameter determined by the equation
\be
\frac{2 \xi}{\pi} \sin(\pi/\xi) = 4 \int_0^{\infty} {\rm d} x \,
\frac{J_0(x) J_1(x)}{x [1 + \exp({U} x/(2\, V_{\textrm{link}}\,))]}
\label{eq_xi},
\ee
and $J_{i=0,1}(x)$ are Bessel functions. A particularly interesting property 
of the BALDA is its discontinuity at half-filling 
\cite{LimaOliveiraCapelle:02}: $v_{\rm xc}(1^+)-v_{\rm xc}(1^-)= 
U - 4 V_{\textrm{link}} \cos(\frac{\pi}{\xi})$. 
For the parameters used in this work (see Fig. \ref{vxcplot}), the discontinuity is
both positive and negative. However, even if the physical gap should be
positive, the results appear not to be affected by this change in sign~\cite{KurthStefanucciKhosraviVerdozziGross:10}. 
New parametrizations that alleviate this issue are currently being developed~\cite{Franca:11}.

The adiabatic approximation implies that
\be
\frac{ \delta v_{\rm xc} [n] (t)}{\delta n (t') } = \delta (t-t')\, f_{\rm xc} (n(t)),
\ee
where $f_{\rm xc}=d v_{\rm xc}(n)/dn$, meaning the $\textrm{\small XC}$-response 
kernel is local in time (and in space). This local and instantaneous
approximation becomes valid for Hubbard systems in the limit of
slowly varying density both in space and in time. These conditions are not
satisfied for the quantum transport system under consideration. 
Despite this fact, reasonable densities were obtained using the BALDA for finite
Hubbard chains~\cite{Verdozzi:08} and it is therefore worthwhile to try the 
approximation for quantum transport phenomena. In the present case this approximation
for $v_{\rm xc}$ is only used on the impurity interacting site, since no interactions in
the leads are present.

For future reference we make a connection with the many-body approach of the previous section.
The fully self-consistent Green function for the whole system (i.e. leads plus impurity) 
satisfies the following equation 
\begin{eqnarray}
\label{impurity_EOM_KS}
G_{ij}(z,z')&=& G^{\textrm{\tiny KS}}_{ij}(z,z') \nonumber\\
 &+& \sum_{kl} \Iid \bar{z} d \bar{z}'G^{\textrm{\tiny KS}}_{ik}(z,\bar{z})[ \Sigma_{kl,{\rm{xc}}}(\bar{z},\bar{z}')\nn\\
&-& \delta(\bar{z},\bar{z}') \delta_{kl} {v}_{k,\rm xc}(\bar{z})]G_{lj}(\bar{z}' ,z').
\end{eqnarray}
Since the exact density is given by
both the KS and the exact Green function, {\em i.e.},
$n_k (z) =-i G_{kk}(z,z^+)=-i G_{kk}^{\textrm{\tiny KS}}(z,z^+)$, it follows that
\bea
\sum_{k} \int_{\mathcal{C}} \textrm{d}\bar{z}\, G_{ik}^{\textrm{\tiny KS}}(z, \bar{z}) v_{k,\rm xc} (\bar{z}) G_{ki}(\bar{z},z )  = \nonumber \\
\sum_{kl}\int_{\mathcal{C}} \textrm{d}\bar{z} \textrm{d}\bar{z}' \,  G_{ik}^{\textrm{\tiny KS}}(z, \bar{z}) \Sigma_{kl,{\rm{xc}}} (\bar{z},\bar{z}') 
G_{li}(\bar{z}',z ),
\label{shamschl}
\eea
where $\Sigma_{\rm xc}$ is the many-body self-energy with the Hartree potential subtracted.
If the self-energy is exact then the corresponding $\textrm{\small XC}$-potential that solves this
Sham-Schl\"uter equation \cite{rvl:96} yields the exact density
of the system. We see that the integral kernel on the left hand side of this equation is nonlocal in space and
time. Hence, the solution of this integral equation for $v_{k,{\rm xc}}$ will in general have values on any site $k$. 
This has been confirmed by recent work of Schenk {\em et al.} \cite{Schenk:11}.
It is important to note that this is true even if the many-body interactions are restricted to the impurity site only.
We therefore make an approximation if we set the $\textrm{\small XC}$-potential to zero in the leads.
We will discuss the validity of this approximation in the results section.

\section{Transport through a weakly coupled correlated site}
\label{results_weak}

We perform many-body and density-functional transport calculations for the Anderson impurity model.
The one-site model is fully specified by  three parameters: the Hubbard interaction (or charging 
energy) $U$, the on-site energy $\varepsilon_0$ and the hopping $V_{\textrm{link}}$ 
connecting the interacting impurity site to leads.
The leads on-site energies are $\varepsilon_L=\varepsilon_R=0$ and 
the hopping in the left and right lead $V_L=V_R=V$. All parameters 
are given in units of the lead hopping $V$.  
For times $t<0$ the contacted system is in equilibrium at zero temperature $(\mu=\ve_{F})$ and 
Fermi energy $\ve_{F}$. A constant 
bias $W_\a$ in lead $\a =$  {\rm{\small (L,R)}} is suddenly switched on at $t=0$
after which the time-dependent observables are calculated.
We only consider weak coupling to the leads, {\em i.e.}, $V_{\textrm{link}} \ll V$, since in this regime the role of 
correlation effects is enhanced. The equilibrium Green
function is obtained as the self-consistent solution of the Dyson equation~\cite{sdvl.2009gw}
for different approximate many-body self-energies. 
In the TDDFT calculations the initial state is obtained by a self-consistent static 
DFT calculation~\cite{StefanucciAlmbladh:04}. For the $\textrm{\small XC}$-potential we use the
modified BALDA defined in Section \ref{tddftsec}.

\subsection{Equilibrium results}
\label{equilibrium}
\begin{figure}[b]
    	\begin{center}
		\includegraphics[width=0.49\textwidth]{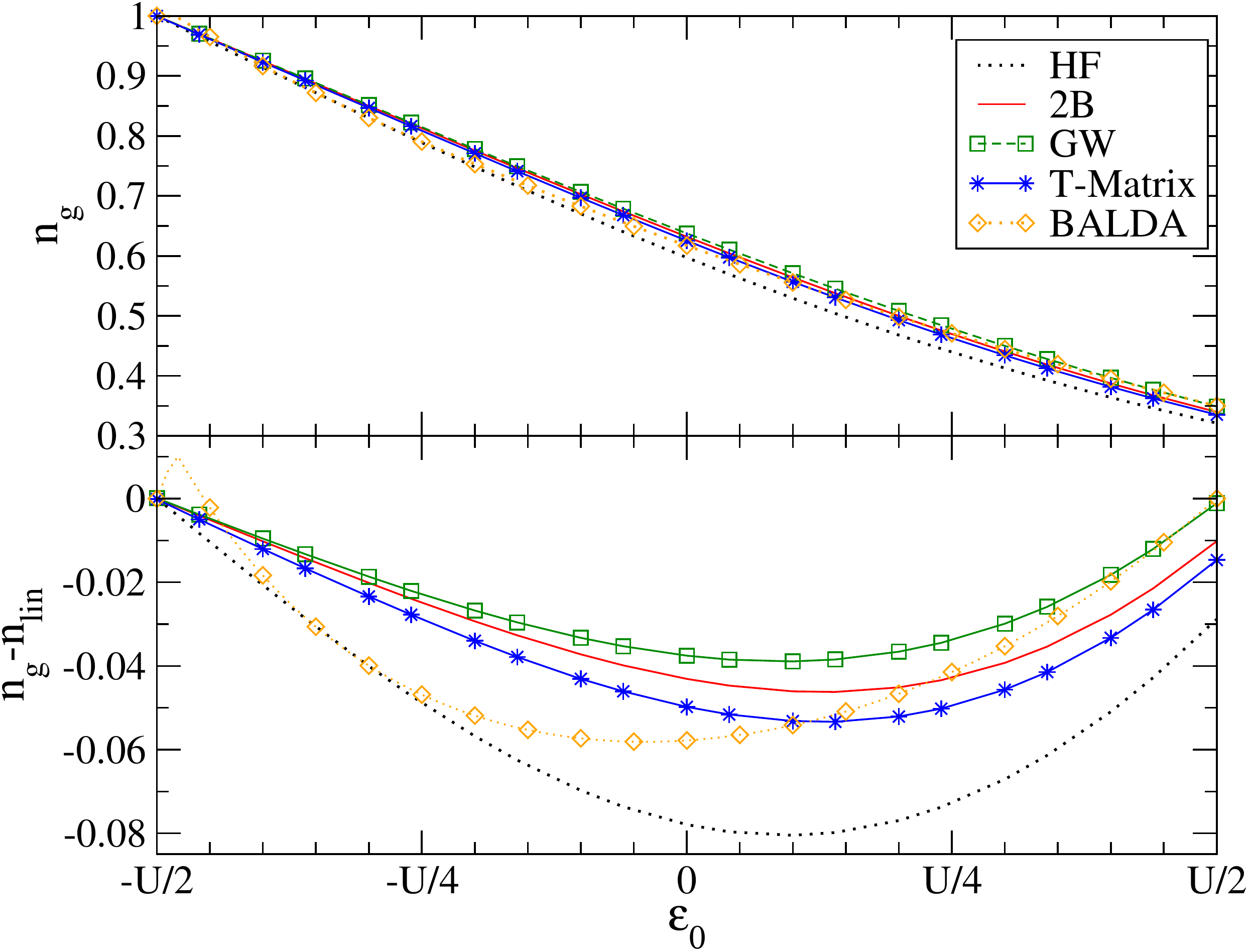}
		\caption{Ground-state density  $n_{g}$ on the 
				correlated site versus the on-site energy, 
				$\varepsilon_{0}$, for $U=1$, $V_{\textrm{link}}=0.5$ 
				and $\ve_{F}=0$. In the bottom panel we subtracted $n_{\rm 
				lin}(\ve_{0})=a\ve_{0}+b$ in order to enhance the difference between 
				the curves. The 
				constants $a$ and $b$ are such that 
				$n_{\rm lin}(-U/2)=1$ and $n_{\rm lin}(U/2)=0.35$.}
	\label{gsdens}
	\end{center}
\end{figure}

We start by considering a system with interaction $U=1$ and coupling to 
leads  $V_{\textrm{link}}=0.5$. The Fermi energy of the
system is $\ve_{F}=0$ (half-filling).
In Fig.~\ref{gsdens} we display the ground-state density $n_{g}$ on the 
correlated site for all values of the on-site energy, $\varepsilon_{0}$, 
for the density functional BALDA and the many-body HF, 2B, GW, and T-Matrix approximations.
For $\ve_{0}=-U/2$ the system is invariant under the particle-hole 
transformation $\hat{d}_{j\s}\ra (-)^{j}\hat{d}^{\dag}_{j\s}$ and therefore 
the exact density on the impurity site equals $n_{g}=1$. 
This remains valid in all the approximation schemes employed.
If we increase the gate potential $\varepsilon_0$ away from the particle-hole symmetric point,
the density on the impurity site decreases almost linearly in all approximations.
In order to enhance the differences between the
approximations, in the bottom panel we plot 
$n(\ve_{0})-n_{\rm lin}(\ve_{0})$ where $n_{\rm 
lin}(\ve_{0})=a\ve_{0}+b$ and the constants $a$ and $b$ are 
chosen such that $n_{\rm lin}(-U/2)=1$ and $n_{\rm lin}(U/2)=0.35$.
In the vicinity of 
the particle-hole symmetric point, the BALDA 
has a cusp that is responsible for correlation induced density fluctuations on
the impurity site. This gives a time-dependent description of the Coulomb 
blockade~\cite{KurthStefanucciKhosraviVerdozziGross:10}. 
The HF approximation can describe the Coulomb blockade provided we allow the
spin symmetry to be broken. The many-body approximations that we use here do not
seem to be able to describe the Coulomb blockade without spin-symmetry breaking~\cite{WangSpataruHybertsenMillis:08}
although the onset of the Coulomb blockade is observed \cite{SchmittAnders:10}.
It can be concluded from the above observations that BALDA yields the Coulomb blockade {\em without} 
spin symmetry breaking~\cite{KurthStefanucciKhosraviVerdozziGross:10}.
For $\ve_{0}<-U/4$ the $\textrm{\small XC}$-potential is close to zero and BALDA consequently
differs substantially from the correlated MBPT results and follows more closely the HF curve. 
When $\ve_{0}$ attains positive values, the correlation potential is large and negative,
favoring charge accumulation (see Fig. \ref{vxcplot}). Consequently, the BALDA deviates 
from HF and follows the correlated MBPT results, in particular with the GW results for $\ve_{0}$
around $U/2$. As a general feature, we find that correlations favor the presence of electrons 
on the interacting site, since the density in the BALDA and the many-body approaches is larger 
than the HF density for all values of the on-site energy.

\subsection{Nonequilibrium steady-state results}
\label{eqss-sec}

\begin{figure}[tbp]
    \begin{center}
	\includegraphics[width=0.5\textwidth]{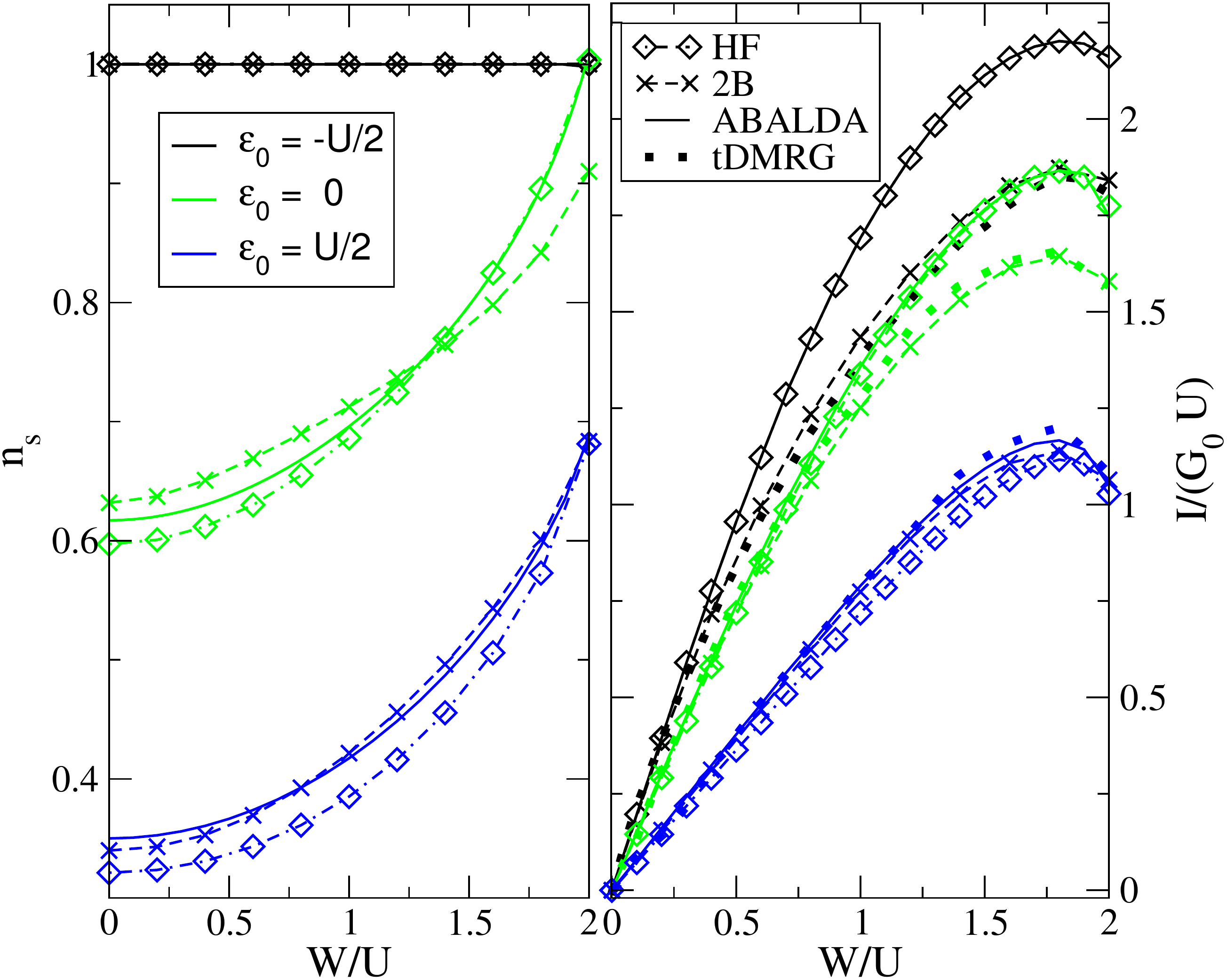}
	\caption{Steady-state density $n_{s}$ (left) and current $I$ (right) 
		for a symmetrically applied bias $W_{L}=-W_{R}=W/2$ and for 
		three different values of the on-site energy 
		$\ve_{0}$. The rest of the parameters are $U=1$, $V_{\textrm{link}}=0.5$ 
		and $\ve_{F}=0$.}
	\label{IV_versus_e}
	\end{center}
\end{figure}

We now shift our attention to the nonequilibrium case. In the left panel of Fig.~\ref{IV_versus_e} 
we display the steady-state density and current (within ABALDA, HF, and 2B) for a symmetrically applied bias 
$W_{L}=-W_{R}=W/2$ and for three different values of the on-site energy 
$\ve_{0}=-U/2\,, 0\,, U/2$. To improve the clarity of the plot we do not display the 
results for GW and T-Matrix as they are, in this parameter range, in close agreement to those 
obtained within 2B. In the left panel of Fig.~\ref{IV_versus_e} we see that the 2B, HF, and ABALDA 
densities are generally in good agreement with each other.

For the corresponding steady-state current, benchmark results are available from tDMRG calculations 
(see Ref.~\onlinecite{HeidrichMeisnerFeiguinDagotto:09}). 
In the right panel of Fig.~\ref{IV_versus_e} we plot the currents as a function of the
bias $W/U$. Because the current is proportional to the overlap of the energy bands of the leads, 
for higher biases, {\it i.e.}, $W/U >1.5$, the steady-state current decrease with increasing bias.

We note that for small bias values all the approximations yield values for the current which 
are on top of the numerically exact tDMRG results for all on-site energies considered. However, 
for higher biases, only the current obtained within 2B follows closely the tDMRG values for all 
on-site energies. Therefore, in this range of parameters, we will use the 2B results for benchmarking 
the other approximations. For $\ve_0=U/2$, the HF and ABALDA results follow closely the tDMRG and 2B curves, 
and for the whole bias-range. For higher biases and smaller on-site 
 energies, {\it i.e.} $\ve_0 =0$ and $\ve_0=-U/2$, they considerably overestimate the exact 
 results. However, the conductances, {\it i.e.} the initial slopes of the I-V curves in 
 Fig.~\ref{IV_versus_e}, still remain in close agreement with the 2B approximation and 
 the tDMRG approach.  This agrees with the Friedel sum rule that relates the conductance 
 to the density~\cite{MeraStefanucci:10}.
 
\begin{figure}[t]
    \begin{center}
	\includegraphics[width=0.5\textwidth]{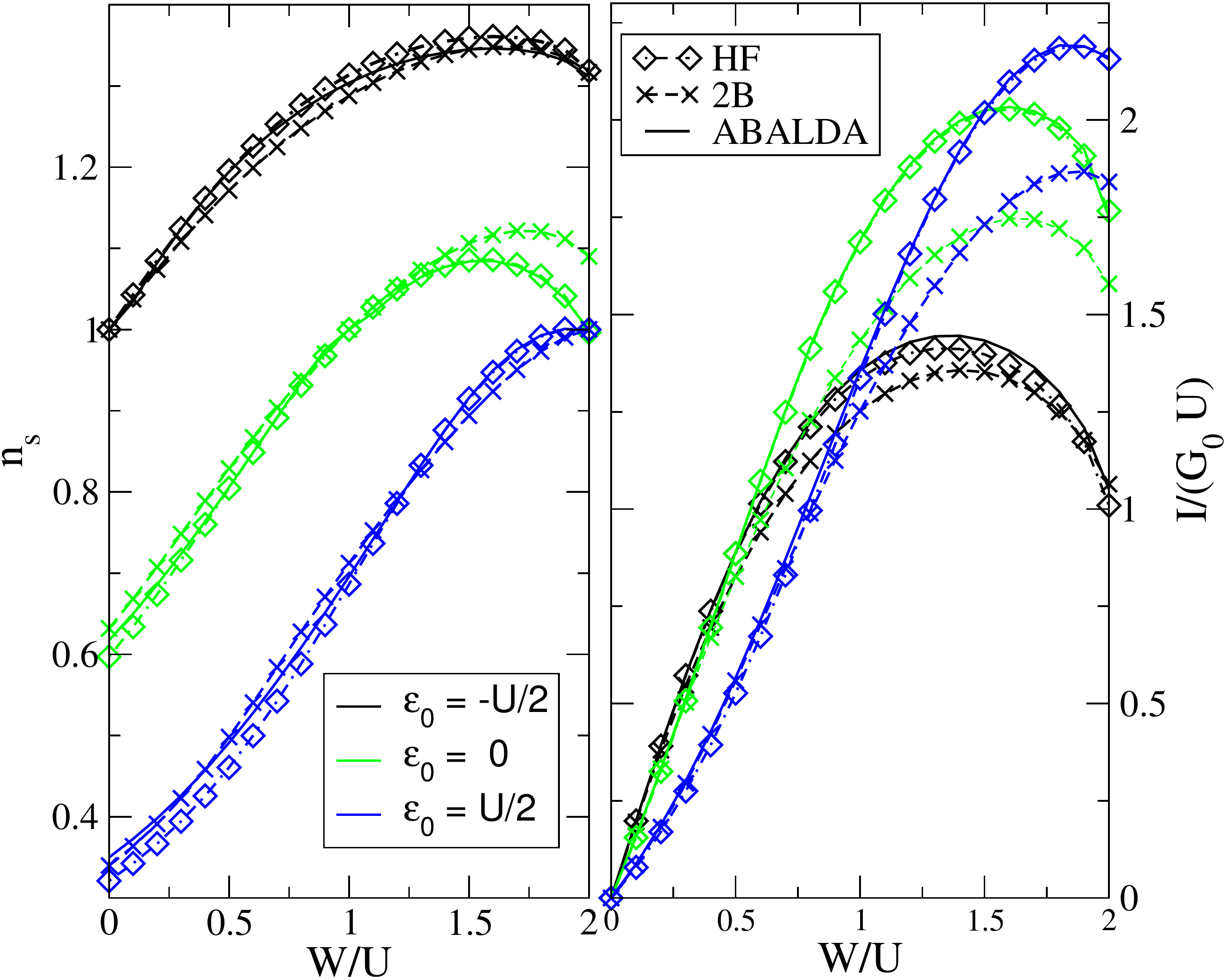}
	\caption{Steady-state density $n_{s}$ (left) and current $I$ (right) 
	for a asymmetrically applied bias $W_{L}=W$, $W_{R}=0$ and for 
	three different values of the on-site energy 
	$\ve_{0}$. The rest of the parameters are $U=1$, $V_{\textrm{link}}=0.5$ 
	and $\ve_{F}=0$.}
	\label{IV_versus_e2}
	\end{center}
\end{figure}

For the results displayed in Fig.~\ref{IV_versus_e2} we considered the same system 
parameters and plotted the density (left panel) and the current (right panel) for an 
asymmetrically applied bias $W_{L}=W$, $W_{R}=0$. The overlap between
the lead energy bands starts to decrease for $W/U>1$ and, consequently, the currents
decrease with increasing bias. The steady-state densities behave similarly to 
the case of symmetric biases (see Fig.~\ref{IV_versus_e}): the ABALDA and the HF 
results are in agreement with 2B results except for the case of gate potential 
$\ve_0=0$ at high bias. For the steady-state current (left panel) we also see the same 
trend: the ABALDA results are close to the HF results and overestimate the 2B results.  
We observe the same trends as in the case of symmetric bias 
(see Fig.~\ref{IV_versus_e}) which indicates that the 2B approximation also here 
gives a description close to the exact result.

\subsection{Time-dependent results: Adiabatic effects}
\label{sec:transient}
We now study the performance of the different approximations in the description of transient phenomena.
The results are compared to the numerically exact tDMRG data of Ref.~\onlinecite{HeidrichMeisnerFeiguinDagotto:09},
obtained for a lead-impurity hopping parameter $V_{\textrm{link}}=0.3535$. This decrease in the hopping parameter 
amounts to a slight enhancement of correlations as compared to the steady-state results of the previous 
section. The tDMRG calculations~\cite{HeidrichMeisnerFeiguinDagotto:09} were done for a particle-hole symmetric 
situation with $\ve_0 = -U/2$. In addition, we compare the many-body results with ABALDA for the on-site 
potential $\ve_0 = U/2$, which is away from the discontinuity of the $v_{\rm  xc}$.

\begin{figure}[t]
    \begin{center}
	\includegraphics[width=0.48\textwidth]{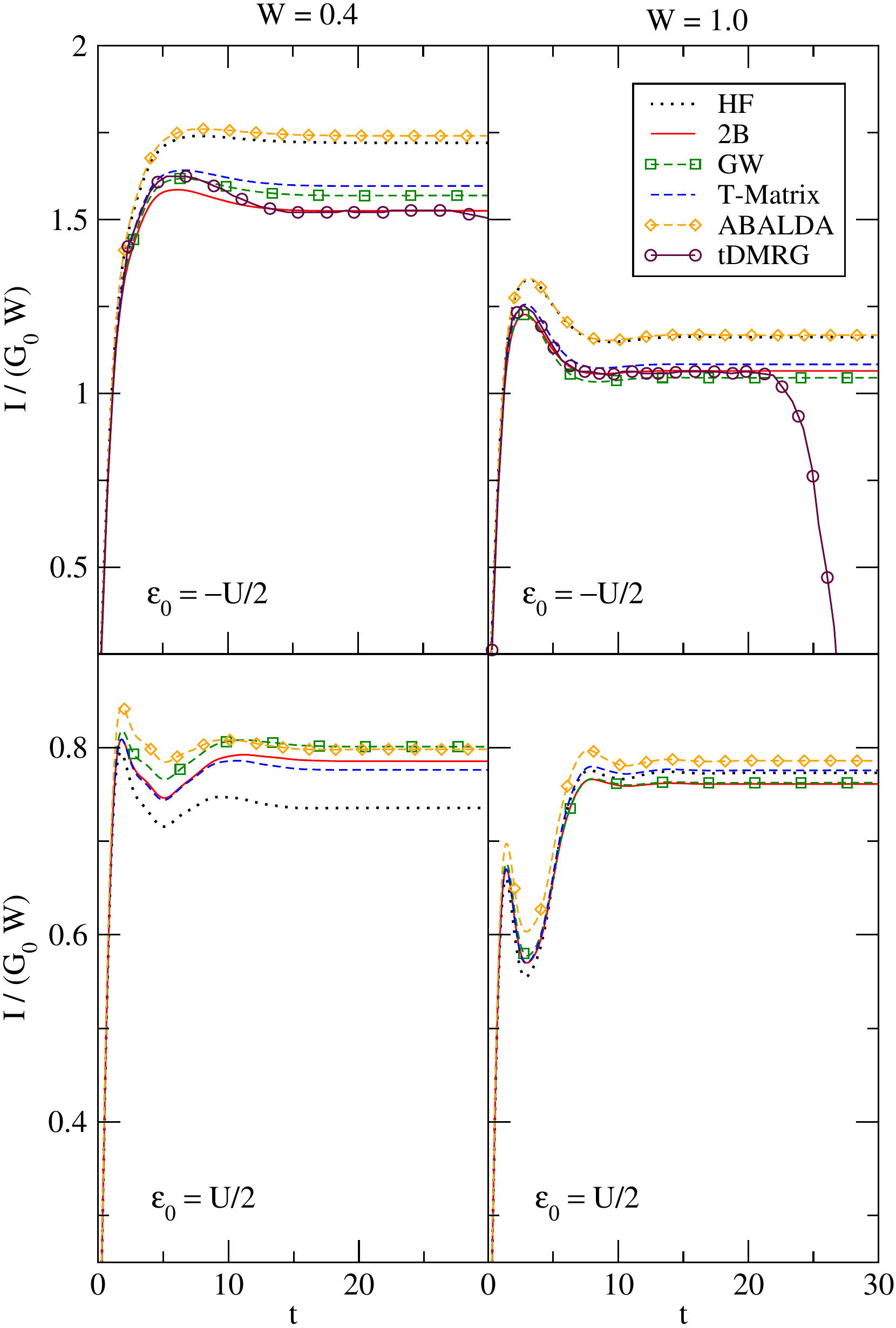}
	\caption{Transient currents for 
			different values of the applied bias $W_{L}=-W_{R}=W/2$,
			$U = 0.5$ and $V_{\textrm{link}}=0.3535$. In the upper panels,
			$\ve_0 = -U/2$ corresponds to the particle-hole 
			symmetric point. In the lower panels $\ve_0 = U/2$.}
	\label{transient}
	\end{center}
\end{figure}

\begin{figure}[t]
    \begin{center}
	\includegraphics[width=0.48\textwidth]{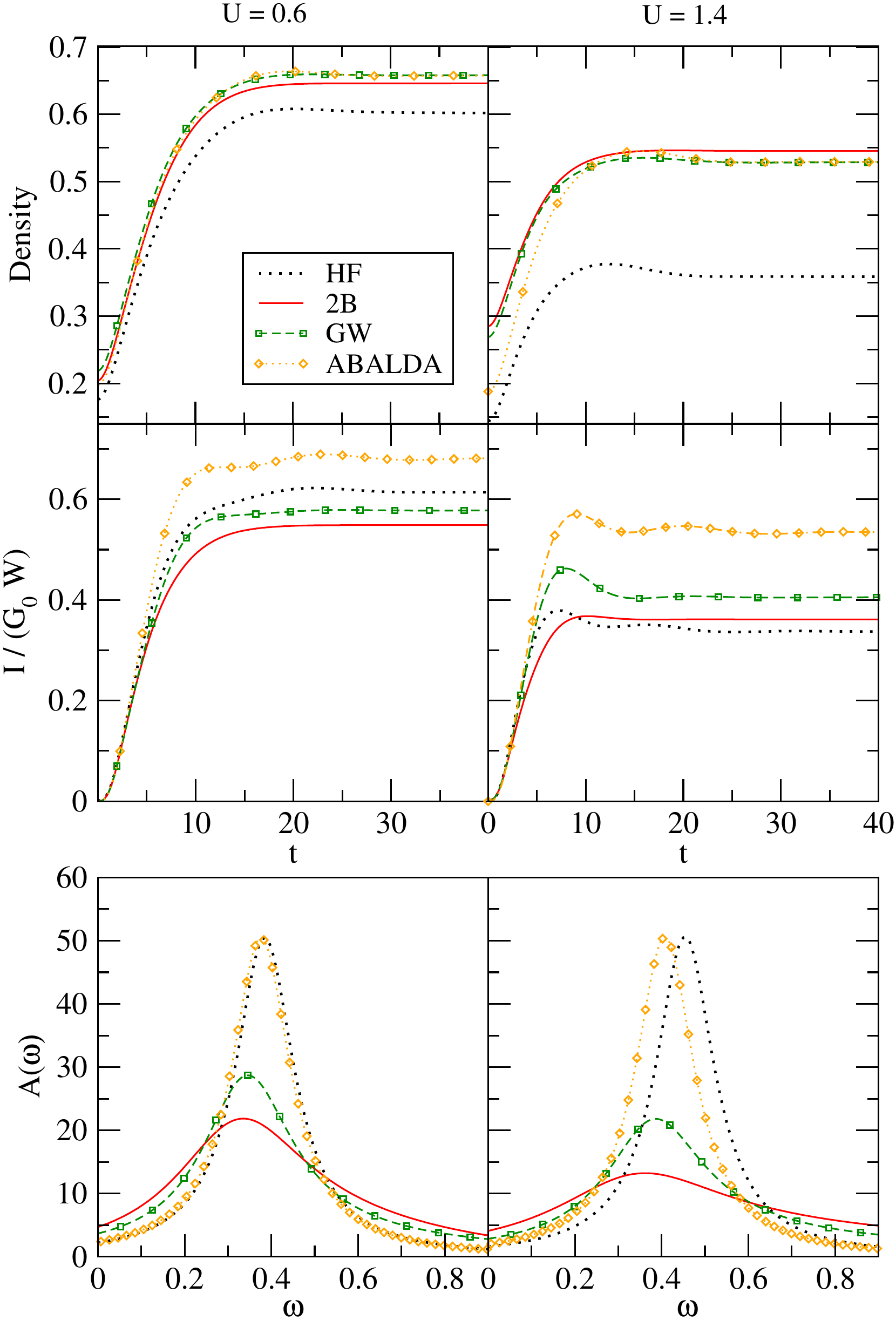}
	\caption{(Color online) Time-dependent density $n_{0}(t)$ for a 
			system with Fermi energy $\ve_{F}=0$, and $\ve_{0}=0.2$, $V_{\textrm{link}}=0.2$ 
			and for different values of the charging energy 
			$U=0.6$ (left column), $1.4$ (right column). The system is driven out of equilibrium by an external 
			bias $W_{L}=0.4$ and $W_{R}=0$. The constant $G_0=e^2/(2\pi \hbar)=1/(2\pi)$ is the quantum of conduction in
			atomic units.}
	\label{fig:balda_comparison}
	\end{center}
\end{figure}

In the upper panels of Fig.~\ref{transient} we display the transient currents as a function of time for the various
many-body approaches and the ABALDA as compared to the benchmark tDMRG data.
Since the tDMRG calculations are performed on finite systems, one sees the influence of reflections at the
system boundaries after a sufficiently long propagation time. The many-body results beyond HF are all in good agreement with the
tDMRG results, the most accurate one being the 2B approximation. Not only the values of the steady-state current
but also the characteristic bump in the transient is well-reproduced. The ABALDA and the HF approximations
perform very similarly; they overestimate the values of the steady-state current and for a bias value of $W=0.4$
they underestimate the height of the transient bump. Also the many-body approximations underestimate 
the height of the bump somewhat. However, the best agreement is again found for the 2B approximation.
It is difficult to pinpoint the origin of the different behavior of the transient bump in the
ABALDA and HF when compared to results obtained within correlated approximations.
It is worth emphasizing, however, that in time-local approximations such as HF the
terms responsible for the initial correlation in the current formula of Eq.~(\ref{eq:current}), 
{\em i.e.}, the terms with components on the vertical track of the Keldysh contour, 
are lacking. In general these terms lead to damping and hence time-local approaches such as
HF tend to overshoot the bump in the transient current \cite{MyohanenStanStefanucciLeeuwen:09}. 
In the upper left panel of Fig.~\ref{transient} such overshoots for the HF and ABALDA are 
probably masked by the fact that the final steady state current goes to a value that is too large.
We finally like to point out that in systems with more levels the transient structure has a more
rich oscillatory time-dependence which can be used to analyze the level structure of the 
central molecule \cite{MyohanenStanStefanucciLeeuwen:09}. In these cases the differences between the HF and 
the correlated approaches become more visible.

In the lower panels of Fig.~\ref{transient}, we display the transient currents the 
on-site energy on the impurity site being $\ve_0 = U/2$.
The transients show a more pronounced oscillatory behavior because of the increased energy-gap between 
the impurity level and the Fermi level of the right lead. This determines the oscillation frequency in the
transient current (see Ref.~\onlinecite{MyohanenStanStefanucciLeeuwen:09}). The many-body approaches agree well
with each other whereas the HF approximation underestimates the value of the steady-state current for lower biases.
In this case the ABALDA results agree closely with the correlated many-body results. Due to the increased on-site energy
$v_{\rm xc}$ becomes negative favoring charge accumulation on the Anderson impurity site (see Fig.~\ref{vxcplot} and 
Section~\ref{equilibrium}).

In order to increase the effects of correlation we now reduce the hopping 
between the interacting site and leads to $V_{\textrm{link}}=0.2$ and
consider two different charging energies $U=0.6$ and $U=1.4$.  
We also set $\ve_{0}=0.2$ and the Fermi energy to $\ve_{F}=0$. 
The system is driven out of equilibrium by a sudden switch-on of a constant, 
asymmetric bias $W_{L}=0.4$, $W_{R}=0$.

In the upper row of Fig. \ref{fig:balda_comparison} we show the time-dependent 
density for the interacting site. For $U=0.6$, all results obtained within correlated 
approximations are in close agreement to each other since if the interaction approaches 
zero, all MBPT approximations become homologous.
As discussed before, in this regime ABALDA and HF are close to the MBPT 
approximations. By increasing the interaction the correlated MBPT approximations
and ABALDA start to detach from HF. 
For stronger interactions, {\em i.e.}, $U=1.4$ (right panel-column), the HF density deviates
considerably from the ABALDA and the many-body results. 

In the middle panel of Fig.~\ref{fig:balda_comparison} we show the time-dependent 
current through the right 
interface (from the interacting site to the lead). As expected from the discussion 
in Section~\ref{eqss-sec} the ABALDA systematically overestimates the current
given by tDMRG and 2B. The deviation from the 2B increases with increasing 
the interaction. The GW approximation also shows a smaller but noticeable 
deviation from the 2B approximation. The agreement between the 
ABALDA and the many-body results deteriorates gradually with an even further 
increase of the charging energy.
For the MBPT results, the differences in the currents when increasing the interaction can be 
explained with the aid of the spectral function. We display the 
{\em steady-state} spectral functions in the lower panel of the Fig.~\ref{fig:balda_comparison}.
Since the current is approximately proportional  
to the integral of the spectral function over the bias window (see Eq.~\eqref{MW}), the highest current 
is given by the approximation which has most spectral weight inside the
bias window. On the other hand, the ABALDA spectral function being very close to the 
HF spectral function does not explain the rather large overestimation of the ABALDA current. 
As in the case of the site densities, for small charging energies 
the spectral functions of all the approximations remain very close 
to each other. 

The spectral functions of correlated MBPT approximations are broadened compared 
to the HF spectral functions. This is because many-body interactions lead to 
a fast decay of many-body states generated by adding and removing particles. 
More precisely, the states 
$| \Psi (t) \rangle = \hat{d}^\dagger_H (t) | \Psi_0 \rangle$ and $| \Phi (t) \rangle = \hat{d}_H (t) | \Psi_0 \rangle$ 
in which we add or remove a particle at 
time $t$ to the impurity in the presence of a bias have decreased survival probabilities 
$| \langle \Psi (t) | \Psi (t') \rangle |^2$ and $| \langle \Phi (t) | \Phi (t') \rangle |^2$ for $|t-t'| \rightarrow \infty$ 
when we include interactions. 
This process is often referred to as quasi-particle scattering. When the charging energy is increased quasi-particle 
scattering broadens the spectral functions and lowers the intensity of the spectral peak in the case of correlated MBPT 
approximations~\cite{ThygesenRubio:08,uimonen:10,MyohanenStanStefanucciLeeuwen:09}. 

The broadening of the HF spectral function is independent of $U$ due to the absence 
of quasi-particle scattering and depends only on the embedding to the leads. 
The same holds true for the ABLADA spectral function which remains very close
to the HF spectral function when increasing the interaction. It should be noted, however, that the 
ABALDA spectral function is the one of the KS system and should not be regarded as an 
approximation to the true spectral function. The clear broadening of the MBPT spectral functions as 
compared to HF demonstrates the importance of non-adiabatic effects in the transient regime. 
Therefore, memory must be 
taken into account for a proper description of ultrafast time-dependent processes.
In the next section we show that memory is, however, not enough to improve the results of the
steady-state current and we identify a second important direction in which to go to improve the
ABALDA.

\subsection{Time-dependent lead densities and non-locality}
\label{sec:lead_dens}

\begin{figure}[t]
	\includegraphics[width=0.49\textwidth]{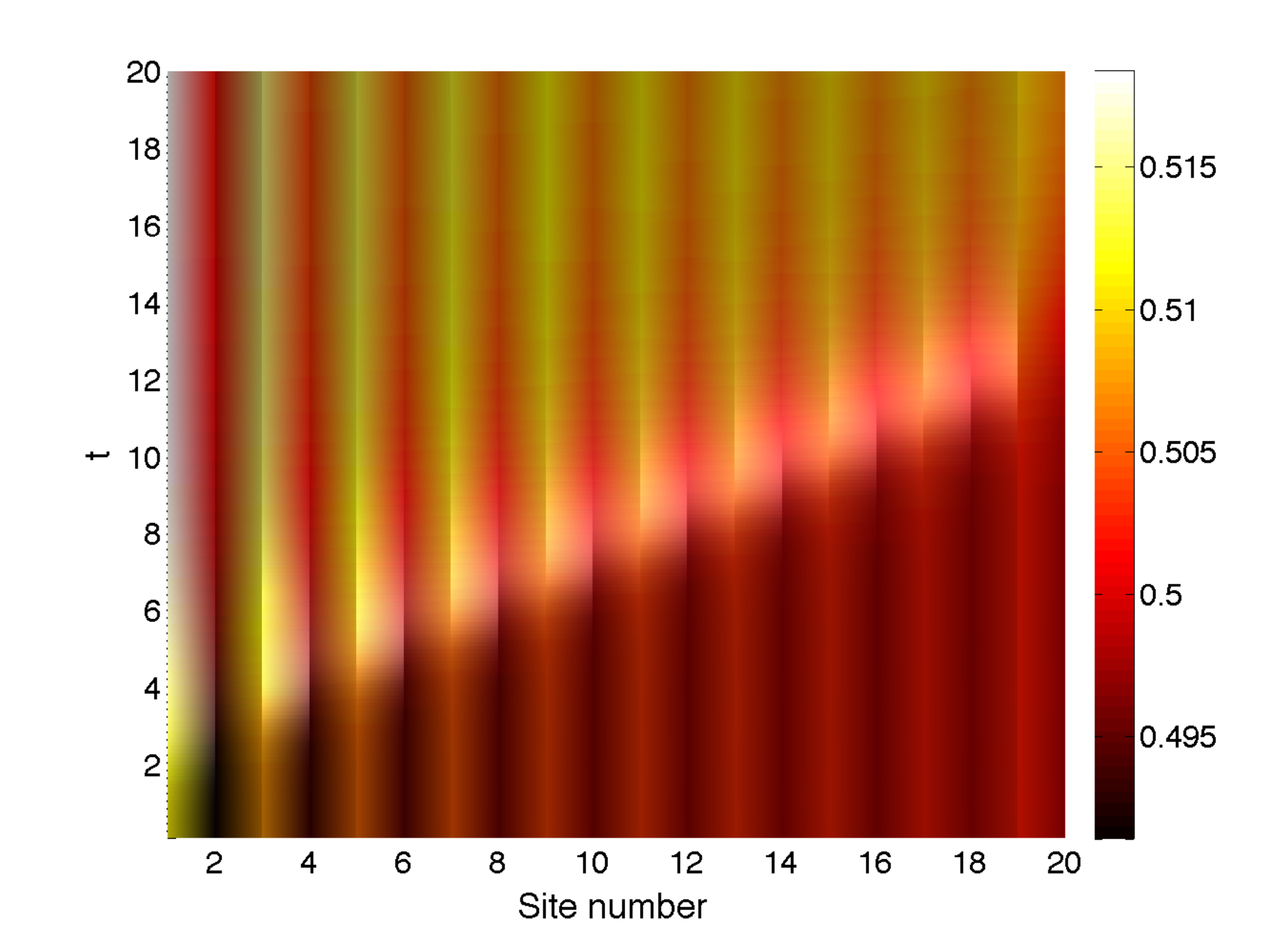}
	\caption{Time-dependent density in the right lead within the 2B approximation for a 
	system with Fermi energy $\ve_{F}=0$, and $\ve_{0}=0.2$, 
	$V_{\textrm{link}}=0.2$ and $U=0.6$). The system is driven out of equilibrium by an external 
	bias $W_{L}=0.4$ and $W_{R}=0$. A density wave entering the lead can clearly be observed. } 
	\label{fig:lead_dens}
\end{figure}

In order to gain some insight on how to cure the deficiencies  of the ABALDA 
$\textrm{\small XC}$-potential, so as to yield an improved time-dependent current, 
we argue as follows:
In equilibrium, the density deep inside the leads is the same in all 
approximations and it is uniquely determined by the Fermi energy $\ve_{F}$. 
Let us denote with $n_{g}$ ($g$=ground state) the density at a site with index $j_{d}$ deep 
inside, say, the right lead, such that $n_{j}=n_{g}$ for all $j>j_{d}$. 
If we plot the current $I_{d}(t)$ to the right of $j_{d}$, no difference will be observed 
in the site-density until after a time $t_{d}=j_{d}/v$, where $v$ is the velocity of
the density wave-front moving into the right lead.
This is clearly illustrated in Fig.~\ref{fig:lead_dens} where we show the time-dependent 
lead densities obtained from a 2B calculation at interaction strenght $U=0.6$ for the first 20 sites in the right lead.
In the lower side of the figure we clearly see a wave front moving into the right lead.

Let us then consider an interval of the right lead that extends from 
$j_{d}$ to $j_{d}+N_{d}$ with $N_{d}\gg 1$. In equilibrium the number of electrons 
in this interval is simply $n_{g}N_{d}$. At the time $t\sim t_{d}$ the current 
wave-front reaches the site $j_{d}$, enters inside the interval 
$(j_{d},j_{d}+N_{d})$ and after a time $T_{d}=N_{d}/v$
it goes out through the site $j_{d}+N_{d}$.

For times $t>t_{d}+T_{d}$ an equal amount of electrons enters in and 
exits from the interval, and a {\em local} steady-state is reached. 
The number of electrons in the considered interval is then given by
\be
n_{s}N_{d}=n_{g}N_{d}+\int_{t_{d}}^{t_{d}+T_{d}}\!\!\!\!dt\, 
I_{d}(t)\sim n_{g}N_{d}+I_{s}T_{d},
\ee
with $I_{s}$ the value of the steady-state current. Taking into 
account that $T_{d}=N_{d}/v$ we conclude that the steady-state 
density deep inside the leads must be
\be
n_{s}=n_{g}+I_{s}/v.
\label{leaddenscurrent}
\ee
\begin{figure}[t]
	\includegraphics[width=0.45\textwidth]{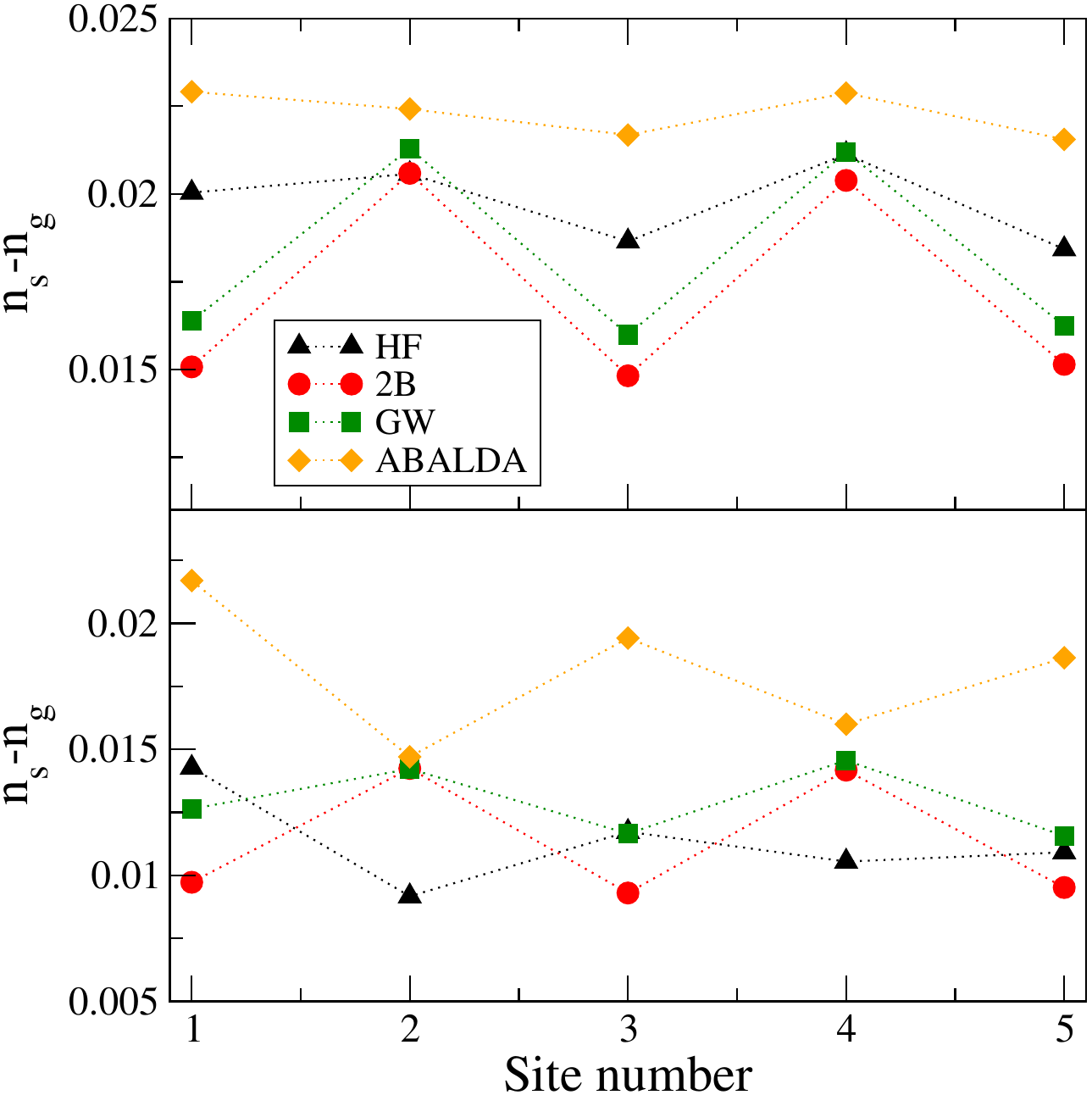}
	\caption{Difference between steady-state and ground-state density in the
	right lead for a system with Fermi energy $\ve_{F}=0$, and $\ve_{0}=0.2$, 
	$V_{\textrm{link}}=0.2$ and for different values of the charging energy 
	$U=0.6$ (top panel) and $U=1.4$ (bottom panel). The system is driven out of equilibrium by an external 
	bias $W_{L}=0.4$ and $W_{R}=0$.}
	\label{nndens}
\end{figure}
From Fig.~\ref{fig:lead_dens} we see that for our 2B calculation the velocity $v$ has the
value $v=1.88$. Given the value of the current of $I_s =0.034$ for this case ($U=0.6$)
we find that the density difference $n_{s}-n_{g}$ is approximately $0.018$ which is in good agreement with the value in the
upper panel of Fig.~\ref{nndens}. 
Also for the case of the $U=1.4$ interaction strength we see from the lower panel of Fig.~\ref{nndens} that the 
ratio of the density differences $n_s-n_g$ for ABALDA and 2B is the same as the corresponding ratio
for the currents in Fig.~\ref{fig:balda_comparison}.
We note that the value $v$ is close to the Fermi
velocity in the lead at half-filling as obtained from a semi-classical calculation. This 
is given by  $v=2 V$. Equation~(\ref{leaddenscurrent}) 
shows that if different approximations yield different values of the 
steady-state current they must also yield different values of the steady-state 
density deep inside the leads.
This can indeed be seen in Fig.~\ref{nndens} where we plot $n_s-n_g$ for the various approximations
for the first five sites in the right lead. The ordering of the density differences is
identical to that of the currents in Fig.~\ref{fig:balda_comparison}. Therefore,
the ABALDA overestimates the difference between steady-state and ground state densities in the
leads. However, ABALDA gives a quite good description of the density on the impurity site, comparable to those
obtained within the many-body approximations. 
We thus conclude that the ABALDA xc-potential is quite accurate on the impurity site but that
setting the potential to zero in the leads is a too crude approximation.
As was discussed in relation to the Sham-Schl\"uter equation (see Eq.~\ref{shamschl})
the $\textrm{\small XC}$-potential will in general have values in the leads even when the
interaction is localized on the impurity site only.
Hence, in order to obtain accurate values for the current within a TDDFT approach
one needs an $\textrm{\small XC}$-potential that has a nonzero value in the leads.
We wish to observe that this nonlocality is different in nature from the non-local
dependence of the $\textrm{\small XC}$-potential on the density. The latter is already
implied by the conclusions of the previous section since non-locality in time and space are
intimately related by conservation laws.

\section{Conclusions and Outlook}
\label{sec:conclus}

We study electron transport through an interacting Anderson impurity model within 
TDDFT and MBPT frameworks. Results obtained in the ground-state, transient and 
steady-state regimes are compared with numerically exact tDMRG values.

In the ground state, we find that for large values of the on-site energy, the density obtained 
using the ABALDA $\textrm{\small XC}$-functional is close to the densities obtained within correlated 
MBPT approximations. However, for smaller values of the on-site energy, the difference 
between the ABALDA and the correlated MBPT densities is significant, ABALDA being closer to HF in this 
parameter range.

In all the cases where benchmark tDMRG results are available we find that the MBPT
approximations beyond HF which we considered give densities and currents 
close to the benchmark ones for the entire parameter range considered. This is true for
both the transient and steady-state regimes. We find that in particular the 2B
approximation performs very well. 
The transients obtained within the 2B approximation are the closest to the tDMRG ones, 
while the HF and ABALDA transients deviate significantly. 
This indicates that it is important to include memory or retardation effects
to properly describe quasi-particle scattering in non-equilbrium transport.

Regarding the TDDFT approach we find that the ABALDA performs very well and yields 
accurate densities on the interacting site but in many 
cases overestimates the steady-state currents.
This problem can be linked to an overestimation of the lead-densities within the 
ABALDA. The results strongly suggest that it is necessary to go beyond the local 
approximation and that one especially needs to take into account XC-potentials 
that are nonlocal and that are non-zero within the leads. Improved functionals 
should therefore be nonlocal functionals in space. As has been clearly pointed 
out by Vignale \cite{VignaleKohn:96}, this implies that the functionals also need to 
be nonlocal in time in order to satisfy basic conservation laws. The construction 
of such functionals is a clear challenge for the future. One way to proceed would 
be to make connections to many-body theory with conserving 
approximations~\cite{vonBarthetal2005}.

\subsection{Acknowledgements}
\label{sec:ack}

Part of the calculations were performed at the CSC -- IT Center for Science Ltd administered by 
the Ministry of Education, Science and Culture, Finland.  Stefan Kurth acknowledges 
funding by the "Grupos Consolidados UPV/EHU del Gobierno Vasco" (IT-319-07) and the 
European Community's Seventh Framework Programme (FP7/2007-2013) under grant 
agreement No. 211956. Adrian Stan acknowledges funding by the Academy of 
Finland under grant no. 140327/2010. We acknowledge the 
support from the European Theoretical Spectroscopy Facility.



\end{document}